\documentclass[aps,superscriptaddress,showpacs,notitlepage,prd]{revtex4-1}
\usepackage{graphicx}
\usepackage{subfigure}
\usepackage{grffile}
\usepackage{natbib}
\usepackage{amssymb}
\usepackage{wasysym}
\usepackage{color}
\newcommand{\beq}{\begin{equation}}
\newcommand{\eeq}{\end{equation}}

\begin{document}

\title{On the stability of black holes with nonlinear electromagnetic fields}

\author{Nora Bret\'on}
\affiliation{Departamento de F\'{\i}sica, Centro de Investigaci\'on y de Estudios Avanzados del I. P. N., Apdo. 14-740, D.F., M\'exico.}

\author{Santiago Esteban Perez Bergliaffa}
\affiliation{Departamento de F\'{\i}sica Te\'orica, Instituto de F\'{\i}sica, Universidade do Estado do Rio de Janeiro, 
Brasil.}

\begin{abstract}
The stability of three static and spherically symmetric black hole solutions with nonlinear electromagnetism as a source is investigated in three different
ways. We show that the specific heat of all the solutions displays an infinite discontinuity with a change of sign , but the turning point method indicates that the solutions are thermodinamically stable (much in the same way as is the case of the Reissner-Nordstrom geometry). We also show that the black holes analyzed here 
are dynamically stable, thus suggesting that there may be a 
relation between thermodynamical and dynamical stability for 
nonvacuum black holes.

\end{abstract}

\pacs{04.70.Bw, 04.20.Jb, 04.40.Nr, 11.10.Lm}

\maketitle

\section{Introduction}

The stability of black hole solutions can be analyzed from 
different points of view. First, since black holes have been considered as thermodynamical systems
after the papers by Bekenstein (who postulated the relation between the area of the horizon and the entropy) \citep{Bekenstein1973}, and Hawking (who showed that they posses a nonzero temperature, due to Hawking radiation) \citep{Hawking1974}, one can wonder about their thermodynamical stability. 
In this regard, Davies \citep{Davies1977} demonstrated that the specific heat (at constant charge or angular momentum) 
of the Kerr-Newman black hole presents an infinite discontinuity along which it changes sign
(while the Gibbs free energy and its first derivative are continuous), and associated this discontinuity to an instability stating that it would lead to a second-order phase transition. The thermodynamical stability of black holes can also be studied using the so-called Poincar\`e (or turning point) method \footnote{See \citep{Poincare1885} for the original version, and 
\citep{Sorkin1982} and \citep{Schiffrin2013}
for updates.}, which asserts that 
changes in stability of a series of equilibrium states can only occur 
when there is a vertical tangent in the plot of conjugate pairs of variables 
(such as mass and the inverse temperature, in the case of a black hole), or when there is a bifurcation.
Using this method, isolated black hole solutions in General Relativity have been shown to be 
thermodynamically
stable in 
\citep{Katz1993, Kaburaki1993}, a result that is at odds with 
the findings of Davies
\footnote{For different points of view about this discrepancy, see
\citep{Sokolowski1980,Sorkin1982, Pavon1988, Pavon1991, Katz1993, Kaburaki1993,Lousto1993, Arcioni2004, Ablu2010,Parentani1995, Okamoto1995}}.
%

Second, several analyses of the stability of black holes from a dynamical point of view have yield the result that black hole solutions of General Relativity in four dimensions are
stable at the linear level.
In the case of Schwarzschild's solution, the stability was proven by Regge and Wheeler \citep{Regge1957}, while that of the Reissner-Nordstrom (RN) geometry was shown in \citep{Moncrief1974a, Moncrief1974b}. The proof of the stability of Kerr's solution
was given in \citep{Whiting1989}. 

The relation between thermodynamical and dynamical stability was recently discussed in \citep{Hollands2012} (see also \citep{Schiffrin2013,Green2013}). In particular, it was shown there that 
for vacuum black holes in General Relativity
dynamical stability is equivalent to thermodynamic stability,
for perturbations with $\delta M = \delta J = \delta A = 0$. 
Hence, a turning point implies in this case a dynamic instability. 
We would like to take here one step further in the understanding of the relation between these different types of stability by 
exploring in detail some exact solutions representing nonvacuum 
black holes in General Relativity.
In particular, 
we will focus on
static and spherically symmetric charged black holes with nonlinear electromagnetism as a source. 
These solutions, which  
generalize the RN geometry,
have received considerable attention recently. The Born-Infeld-Einstein
static and spherically symmetric (SSS) spacetime was presented in \citep{Garcia1984, Breton2003}, 
and its thermodynamical properties analyzed in \citep{Chemissany2008, Gunasekaran2012}. A regular black hole geometry in the presence of a nonlinear electromagnetic field
which reduces to that of Maxwell in the weak-field limit
was obtained in \citep{Ayon1998} using the dual formalism introduced in 
\citep{Salazar1987}, and further discussed in \citep{Baldovin2000} and \citep{Bronnikov2000}. The SSS black hole with the Euler-Heisenberg effective
Lagrangian of quantum electrodynamics as a source was examined in \citep{Yajima2000}, and
the same type of solutions with Lagrangian densities that are
powers of Maxwell's Lagrangian were analyzed in \citep{Hassaine2008}. 
Also worth mentioning are the general analysis of \citep{Diaz2009, Diaz2010}, the examination of the thermodynamics of black holes 
with an arbitrary nonlinear Lagrangian for the electromagnetic field 
in \citep{Rasheed1997} and of the Smarr formula in \citep{Breton2004},
and the enhanced no-hair conjecture presented in \citep{Garcia2012}. 
To close this incomplete list, we would like to point to the references
\citep{Moreno2002} and \citep{Breton2005}, where 
the dynamical stability of SSS black holes with a nonlinear electromagnetic source was examined, and 
\citep{Flachi2012}, where a study of the quasinormal modes of neutral
and charged scalar field perturbations on regular nonlinear electromagnetic black hole backgrounds was presented.

Our main goal is to compare the results of the three abovementioned ways of determining the stability in the case of SSS black holes with NLEM as a source. We shall consider both singular and regular solutions. In particular, we shall see that in all cases there is a discontinuity in the specific heat of the kind present in the RN solution, but no sign of instability according to the Poincar\`e method.
As a byproduct, we will exhibit several expressions 
valid for any static and spherically symmetric charged black hole, and a study of the position of the horizons in the 
exact solutions under scrutiny. 

\section{Background}
\label{backg}
Several equations that follow from the assumed symmetries of the problem will be deduced in this section. 
The action
for the gravitational and the electromagnetic field is
given by (the notation used in this paper agrees with that in \citep{Bronnikov2000})
\begin{equation}
S= \frac{1}{16 \pi} \int{d^4x \sqrt{-g}\;[R-{\cal L}(F)]},
\label{action}
\end{equation}
where $R$ is the scalar curvature, $F_{\mu \nu} \equiv \partial_{\mu}A_{\nu}-\partial_{\nu}A_{\mu}$ is the electromagnetic field tensor, and ${\cal L}$ is an arbitrary function of $F \equiv F^{\mu \nu} F_{\mu \nu}$. 
The equations of motion that follow from Eqn.(\ref{action}) are
\begin{equation}
G_{\mu \nu}=-T_{\mu \nu}
\label{einstein}
\end{equation}
\begin{equation}
\nabla_{\mu}({\cal L}_{F}F^{\mu \nu})=0,\;\;\;\; \quad \nabla_{\mu}(\ast F^{\mu \nu})=0,
\label{maxwell}
\end{equation}
where the subindex $F$ means derivative w.r.t. $F$, 
$\ast F_{\mu \nu}$ is the dual of $F_{\mu \nu}$,
and $T_{\mu \nu}= -2{\cal L}_{F}F_{\mu \alpha}F_{\nu}^{\cdot \alpha}+ \frac{1}{2}g_{\mu \nu} {\cal L} $. From the \textit{Ansatz}
$$
ds^2 = e^{2\gamma (r)}dt^2 - e^{2\lambda (r)}dr^2 - r^2d\Omega^2,
$$
and Eqns.(\ref{einstein}), it follows that $\gamma = - \lambda$. Hence we adopt the notation
\begin{equation}
ds^2=\left(1- \frac{2{\cal M}(r)}{r}\right)dt^2-\left(1- \frac{2{\cal M}(r)}{r}\right)^{-1}dr^2-r^2 d\Omega^2.
\label{metric}
\end{equation}
Einstein's equations also yield
\begin{equation}
{\cal M}(r) =  k +\frac{1}{2} \int_{r_h}^{r}{T^{0}_{0}(r) r^2dr}.
\label{m}
\end{equation}
where $k$ is an integration constant.
The electromagnetic tensor compatible with spherical symmetry has two nonzero components ($F_{01}=-F_{10}$ and $F_{23}=-F_{32}$), corresponding to the radial electric and magnetic fields. In each case, it follows from Eqns.(\ref{maxwell}) that
\begin{equation}
r^2{\cal L}_{F}F^{01}=Q_e, \;\;\;\;\;\;\;\;\;\;\;\;\;\;\;\;\;\;\;\;\; F_{23}=Q_m \sin \theta,
\end{equation}
where $Q_e$ and $Q_m$ are the electric and magnetic charges, respectively.
With the definitions 
$$f_e \equiv 2F_{01}F^{10}=2Q_e^2{\cal L}_{F}^{-2}r^{-4} \ge 0, \;\;\;\;\;\;\quad f_m=2F_{23}F^{23}=2Q_m^2 r^{-4} \ge 0,
$$
the energy-momentum tensor can be written as
\begin{eqnarray}
T^\mu_\nu
&=& \frac{1}{2}\; {\rm diag} \left({\cal L}+2f_e{\cal L}_{F},{\cal L}+2f_e{\cal L}_{F}, {\cal L}-2f_m{\cal L}_{F}, {\cal L}-2f_m
{\cal L}_{F}\right).
\label{EE}
\end{eqnarray}
It follows that $T^0_{\;0} =  T^1_{\;1}$, 
and $T^2_{\;2} = T^3_{\;3}$.
The conservation of the energy-momentum tensor, $\nabla_{\mu}T^{\mu \nu}=0$,
yields
\begin{equation}
2T^{1}_{1}+r(T^{\;\prime\;1}_{1})-2 T^{2}_{2}=0,
\label{ec}
\end{equation} 
where a prime denotes the derivative wrt $r$. 
Combining the previous equation with $R=T = 2T^0_0 + 2 T^2_2$ we find that 
\begin{equation}
R = r(T^{\prime\;0}_{0})+ 4 T^0_0.
\end{equation}
These expressions will be used in the forthcoming sections, as well as  
in the calculation of the specific heat for an SSS black hole with a nonlinear electromagnetic field as a source, which we present next.
The thermal capacities of 
a black hole
are given by 
$$
C_X = \frac{\kappa}{4}\left(\frac {dA}{d\kappa} \right)_X,
$$
where $X$ is a set of parameters that are being held constant, $\kappa$ is the surface gravity, and $A$ the area, and all the functions are evaluated 
at the horizon radius $r_h$. 
The usual way to calculate $C_X$ involves the use of
the Smarr formula
\citep{Davies1977}.
Since in the case of black holes with a nonlinear electromagnetic source
this formula is no longer valid \citep{Rasheed1997, Breton2004}, we shall follow a different route.  
For SSS black holes, the surface gravity $\kappa$ is given by  
$\kappa = \frac 1 2\left.\frac{d{g_{00}}}{dr}\right|_{r_h}$. Taking into account that  
$g_{00}= 1-2{\cal M}(r)/r$, and ${\cal M}(r)$ is given by Eqn.(\ref{m}),
%
it follows that 
\begin{equation}
\kappa = \frac{1}{2r_h}\left[1-r_h^2T^0_{\;0}(r_h)\right],
\label{kappa}
\end{equation}
an expression which agrees with
the one obtained in \citep{Visser1992}.
The
specific heat at constant charge, defined by 
$C_Q = \frac{\kappa}{4}\left(\frac {dA}{d\kappa} \right)_Q
$
can be calculated 
using
\begin{equation}
C_Q = \frac{\kappa}{4}\left(\frac {dA}{dM}\frac {dM}{d\kappa} \right)_Q,
\end{equation}
with $\kappa = \kappa (r_h)$ given by Eqn.(\ref{kappa}), $A=4\pi r_h^2$,
and  
$r_h=r_h(M,Q)$, and $M$ and $Q$ are the total mass and charge
of the black hole \footnote{The calculation of $C_\Phi$, and of the analogous of the thermal expansion and the isotermal compressibility involve Legendre transformations, and the use of the Smarr formula, which is not available in this context.}.
A straightforward calculation (which is independent of the value of the integration constant $k$ in Eqn.(\ref{m}))
yields
\begin{equation}
C_Q=-2 \pi r_{h}^{2}\; \frac{1-r_{h}^{2} T^{0}_{\;0}(r_h)}{1+r_{h}^{2}T^{0}_{\;0}(r_h)+r_{h}^{3} T^{\prime\; 0}_{\;0}(r_h)},
\label{cq}
\end{equation}
where a prime denotes the derivative with respect to the radial coordinate.
This expression is actually valid for any charged SSS black hole, 
and it reduces to the well-known formula
for the case
of the RN black hole, given by
\begin{equation}
C^{\rm{(RN)}}_Q=-2 \pi r_{h+}^2 \frac{r_{h+}^2-Q^2}{r_{h+}^2-3Q^2},
\end{equation}
where $r_{h+}$ is the external horizon of the RN solution.
The correct expression for the Schwarzschild solution follows from the latter when $Q=0$.

Defining $D = -(1+r_{h}^{2}T^{0}_{\;0}(r_h)+r_{h}^{3} T^{\prime\; 0}_{\;0}(r_h))$, 
and taking into account that \citep{Bekenstein1998}
\beq
0<T_0^0(r_h) \le 1/r_h^2,
\label{desbek}
\eeq
it follows that
the sign of $C_Q$ 
is given by that of $D$, and any possible divergence in $C_Q$ must arise
from a zero of $D$. In fact, using Eqns.(\ref{ec}) and (\ref{desbek}), it follows that
a necessary condition for $D=0$ is  
$$
\frac{1}{2r_h^2} < \left.(T^0_0-T^2_2)\right|_{r_h}\leq \frac{1}{r_h^2}.
$$
For completeness, we also give the expression of 
$C_Q$ in terms of ${\cal M}(r)$ and its derivatives evaluated at $r_h$, 
\begin{equation}
C_Q=-2 \pi r_h^2 \frac{1-2{\cal M}'(r_h)}{1+2{\cal M}'(r_h)+2r_h{\cal M}''(r_h)}.
\end{equation}
We shall present next the plots of $C_Q$ for several analytic SSS black hole solutions corresponding to different NLEM theories, along with an analysis of the position of the external horizon in each case. 

\subsection{Born-Infeld black hole}

\noindent The properties of this singular solution have been discussed in several articles (see for instance \citep{Garcia1984, Chemissany2008,Breton2002,Breton2003}). The Born-Infeld Lagrangian is given by
\begin{equation}
{\cal L}=4b^2\left(-1+\sqrt{1+\frac{F}{2b^2}}\right).
\end{equation}  
With the metric element given by Eq.(\ref{metric}), 
the black hole geometry is determined by the function \citep{Garcia1984}
\begin{equation}
g_{00}(R)=1-\frac{2m}{R}+\frac 2 3 R^2 \left(1-\sqrt{1+\frac{q^2}{R^4}}\right)+\frac 4 3 
\frac{q^2}{R}\int_{R}^{\infty}\frac{dz}{\sqrt{z^4+q^2}},
\label{abi}
\end{equation}
where $R\equiv br$, $q\equiv bQ$, and $m\equiv Mb$, and $b$ has units of $\rm [length]^{-1}$.

\begin{figure}
\centering
\mbox{\subfigure{\includegraphics[width=2.3in]{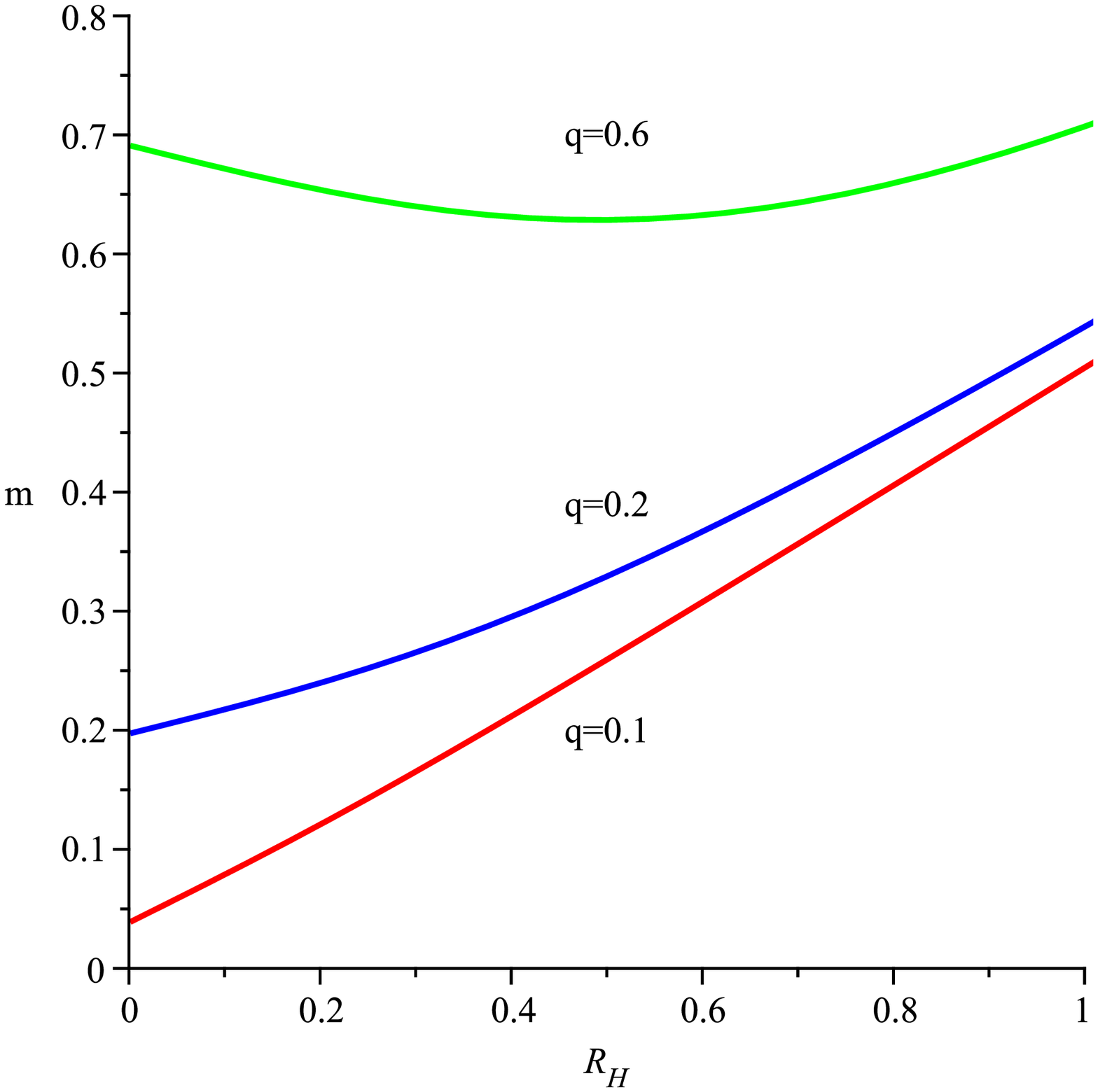}
\quad
\subfigure{\includegraphics[width=2.3in]{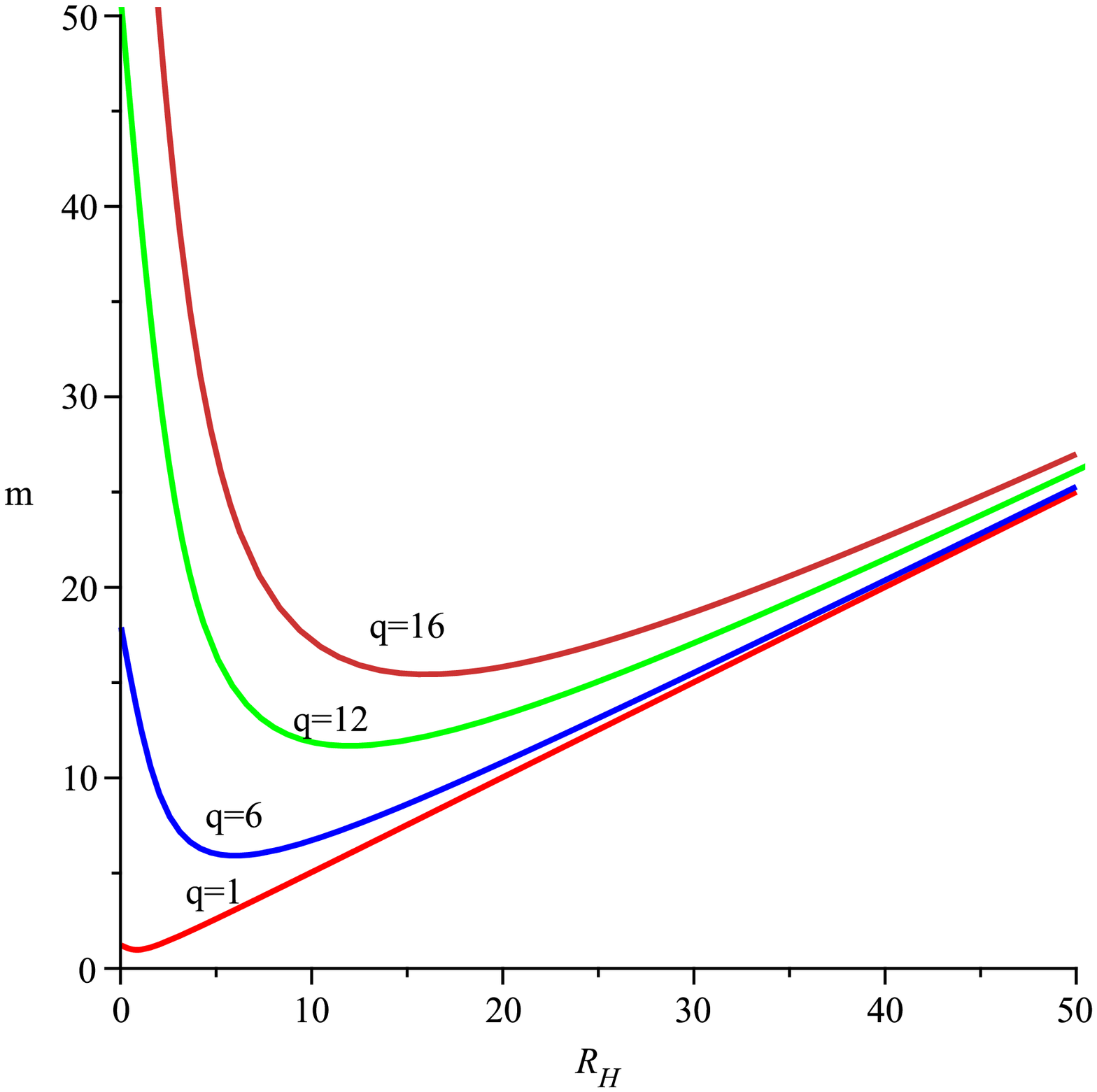}}}}
\caption{The mass of the Born-Infeld black hole as a function of $R_h$ for different values of the charge.}
\label{fig13}
\end{figure}

In order to use Eqn.(\ref{cq}) to calculate $C_q$, the radius of the 
(external) horizon is needed. This is the value $R_h$ such that 
$g_{00}(R_h)=0$. The plots in Fig.\ref{fig13}
show the values of $R_h$ for different values of the parameters. 
Depending on the value of the mass, for
$q < q_{\rm min} = 0.5$, there may be one or no horizon, 
while for $q\geq q_{\rm min}$ there may be two, one, or no horizons
\footnote{The extremal case, in which the two horizons coalesce into one, is given by 
$q^2=\frac{1}{4}+R_h^2$ \citep{Chemissany2008}.}. 
The plots also show that for $R_h>>q$, $R_h \approx 2m$, as can be seen from Eq.(\ref{abi}).

\begin{figure}
\centering
\mbox{\subfigure{
\quad
\subfigure{\includegraphics[width=2.8in]{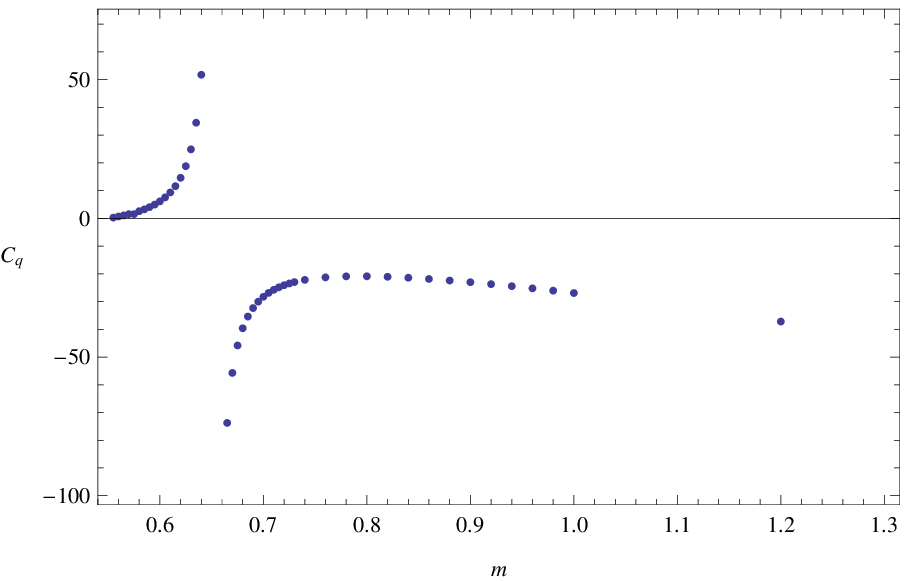}
\quad
\subfigure{\includegraphics[width=2.8in]{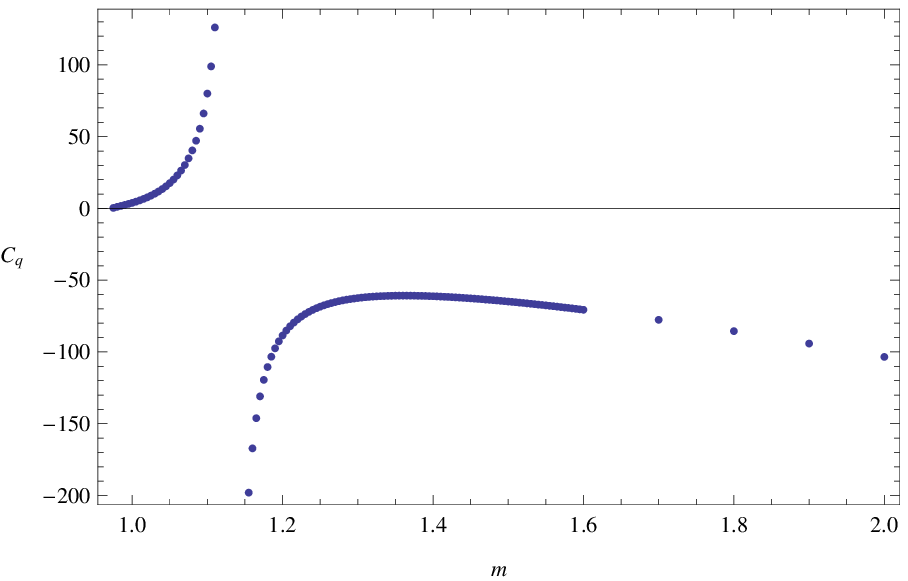}}}}}
\caption{Plot of $C_q$ as a function of $m$ with $q=0.6$ (left) and $q=1$ 
(right) for the Born-Infeld black-hole.}
\label{cqbi}
\end{figure}

The expression for $C_q$ 
for the electrically charged Born-Infeld black hole 
can be computed from Eqn.(\ref{cq}). The 
result is
\begin{equation}
C_{q}=-2 \pi R_{h}^2 \frac{(1+2R_{h}^2)\sqrt{R_{h}^4+q^2}-2(R_{h}^4+q^2)}{(1-2R_{h}^2)\sqrt{R_{h}^4+q^2}+2(R_{h}^4-q^2)}.
\label{ecqbi}
\end{equation}
Fig.\ref{cqbi} shows the plots of $C_q$ as a function of $m$ for two different values of $q$. 
They show a divergence that resembles that of the RN black hole, and reduce to the Schwarzschild's case in the limit $R_h>>q$, as can be checked from Eqn.(\ref{ecqbi}).

\subsection{Nonsingular solutions}

In this section, we shall consider two solutions that are different from the one analyzed in the previous section at least in two respects: (a) they are non-singular, and 
(b) the parameter which measures the gravitational mass 
of the point sources of the electromagnetic field (namely, the integration constant $k$ in Eq.(\ref{m})) is zero, in spite of the fact that these geometries approach asymptotically a RN-like solution. This latter fact is due to the emergence of 
an effective electromagnetic mass $M_{EM}$, defined  in  terms  of  the  asymptotic  behavior  of  the related gravitation field (see for instance
\citep{Pellicer1969}) in such a way that, as we shall see below, $M_{EM}\propto Q^{3/2}$. 

\subsubsection{Bronnikov solution}

\begin{figure}
\includegraphics[width=3in]{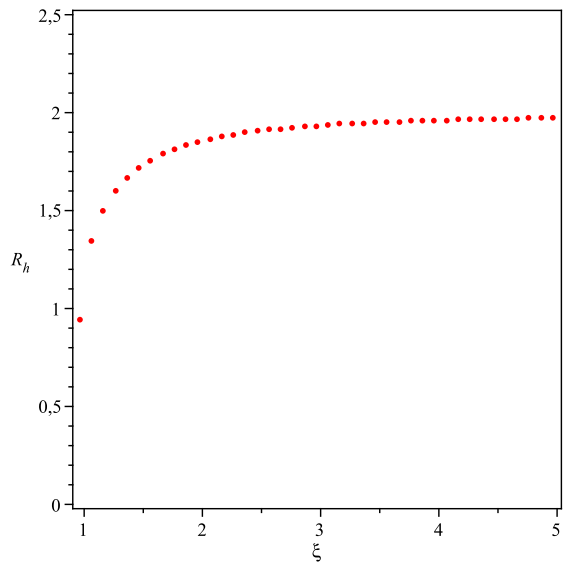}
\caption{Size of the external horizon $R_h$ of Bronnikov's solution in terms of $\xi$, determined by Eqn.(\ref{zerosb})
.}
\label{horbron}
\end{figure}

\begin{figure}[h]
       \centering  
       \includegraphics[width=3in]{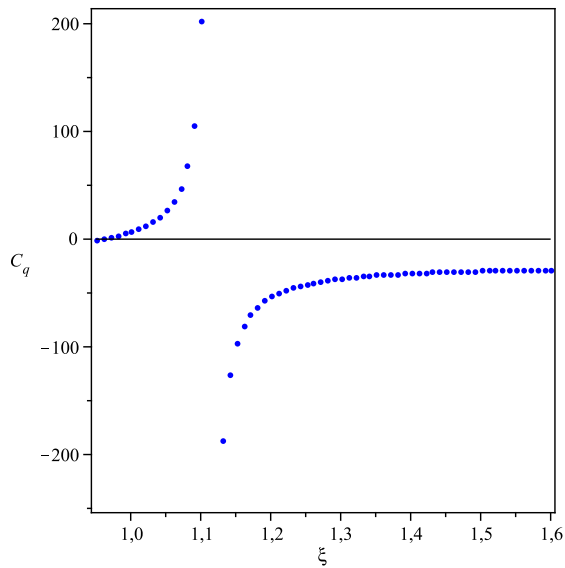}
       \caption{Plot of $C_q$ as a function of $\xi$ for the black hole represented by Eqn.(\ref{g00bron}).}
       \label{cqbr}
\end{figure}

This is a static, spherically symmetric and nonsingular 
magnetic 
black hole solution derived in \citep{Bronnikov2000}, for a theory with Lagrangian given by
\begin{eqnarray}
{\cal L}(F)&=& F {\rm sech}^2 \left[{a\left({\frac{F}{2}}\right)^{1/4}}\right], 
\end{eqnarray}
where $a$ is a parameter \footnote{This Lagrangian was inspired in the one presented in \citep{Ayon1999}.}.
The
relevant metric function is given by
\begin{equation}
g_{00}(r) =  1- \frac{Q_m}{a^2}\left({\frac{a \sqrt{Q_m}}{r}}\right) \left[{1- \tanh{\left({\frac{a \sqrt{Q_m}}{r}}\right)}}\right],
\label{g00bron}
\end{equation}
and the effective electromagnetic mass for this geometry is
\begin{equation}
M_{EM}= M(r \to \infty)=\frac{Q_m^{3/2}}{2a}.
\end{equation}
%
%
The solution behaves as a black hole or as a soliton depending on the value of the parameter $\xi=\frac{M_{EM}}{Q_m}$.
If $\xi=\xi_0\approx 0.96$, then $g_{00}(r)=0$ has only one (double) root, in which case the black hole would be extreme.
If $\xi > \xi_0$, $g_{00}(r)=0$ has two different real roots, hence the black hole has two horizons. Otherwise the solution represents a soliton ($g_{00}(r)>0$ for all values of the radial coordinate). 
The position of the zeros of Eqn.(\ref{g00bron}) is given by \footnote{As shown in \citep{Matyjasek2000}, the location of the horizons can be expressed in terms of the Lambert function.} 
\beq
1-\tanh \left(\frac{1}{2\xi^2R_h}\right) = \frac{R_h}{2},
\label{zerosb}
\eeq
where $R_h\equiv r_h/M$.
The radius of the external horizon in terms of $\xi$ is shown in Fig.\ref{horbron}. It tends to $2M$ for large values of $\xi$.
The expression for the specific heat follows from Eqn.(\ref{cq}):  
\begin{equation}
\frac{C_{Q}}{M^2}=-4\pi R_h^3\xi^2\frac{4R_h\xi^2+R_h-4}{8R_h^2\xi^4-(4-R_h)(6R_h\xi^2+R_h-2)},
\label{cqbron}
\end{equation}
and is plotted as a function of $\xi$ in 
Fig.\ref{cqbr}.  
The plot displays a divergence similar to that of the RN solution, 
and we have verified that it tends to $C_Q^{\rm{(Schw)}}$ for large $\xi$, 
as can be seen from Eqn.(\ref{cqbron}).

\subsection{Dymnikova solution}

\begin{figure}
\centering
\includegraphics[width=3in]{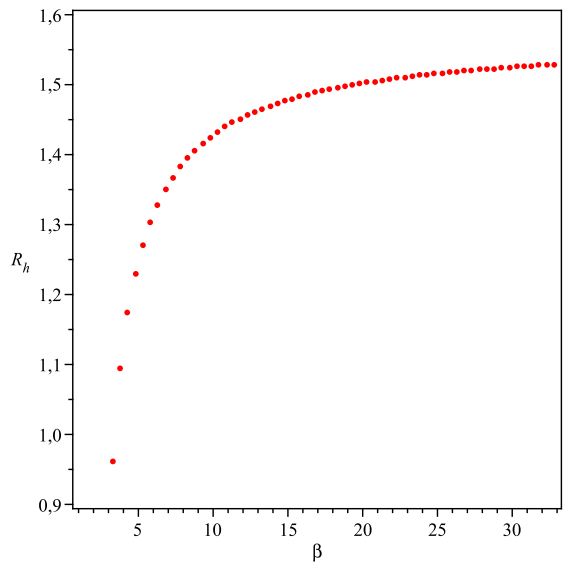}
\caption{Plot of the external horizon as a function of $\beta$ for the solution given in Eqn.(\ref{metdymn}).}
\label{hordymn}
\end{figure}

\begin{figure}[h]
       \centering  
       \includegraphics[width=3in]{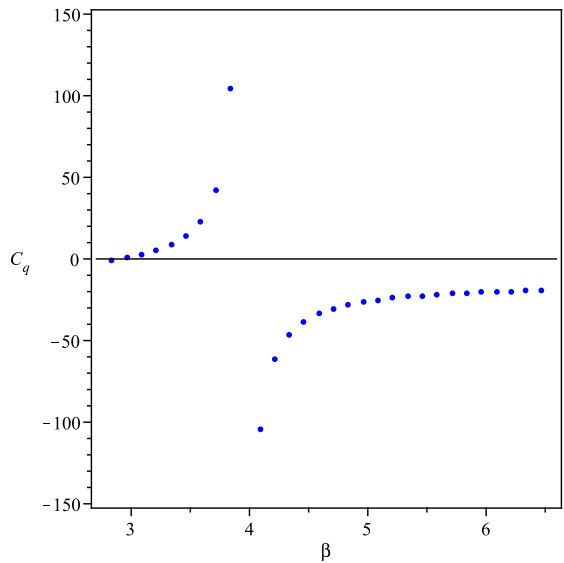}
       \caption{Plot of the heat capacity at constant $Q$ as a function of $\beta$ for the black hole described by Eqn.(\ref{metdymn}).}
       \label{cqd}
\end{figure}

The regular, static and spherically symmetric geometry studied in \citep{Dymnikova2004} has a nonzero electric field 
described by the Lagrangian
\begin{equation}
{\cal L} = \frac{F}{(1+\alpha\sqrt F)^2},
\end{equation}
where $F=-2Q_e^2r^8/(r^2+r_e^2)^6$, $Q_e$ is the electric charge, 
$\alpha = r_e^2/(\sqrt{2}Q_e)$, and $r_e = \pi Q_e^2/(8M)$ is proportional to the classical electromagnetic radius. The metric is determined   by
\begin{equation}
g_{00}(r) = 1-\frac{4M}{\pi r}\left[\arctan\left(\frac{r}{r_e}\right)-\frac{rr_e}{r^2+r_e^2}\right],
\label{metdymn}
\end{equation}
and it goes to a RN-like metric at infinity, as in the previous case. 
The parameter $\beta$ given by
\beq
\beta = \frac{8}{\pi^2}\left(\frac{2M}{Q_e}\right)^2
\eeq
discriminates between a 
regular electrically charged black hole and
a self-gravitating particle-like structure. 
A single-horizon black hole is described by 
$\beta = \beta_0 \approx 2.82$, while the solution 
with two horizons has $\beta > \beta_0$ \footnote{This solution evades the no-go theorem presented in \citep{Bronnikov2000} because the Lagrangian does not go
to the Maxwell's limit at the center,
see \citep{Dymnikova2004}.}.
The position of the horizon(s) is given by 
\begin{equation}
\arctan (R_h\beta) - \frac{R_h\beta}{\beta^2R_h^2+1} = R_h,
\end{equation}
where $R_h \equiv r_h/\mu$, and $\mu \equiv 4M/\pi$.
The variation of the radius of the external horizon 
with $\beta$ is plotted in 
Fig.\ref{hordymn}. The plot shows that the solution tends to the Scharzschild 
black hole for large $\beta$.
The expression for the specific heat is in this case
\begin{equation}
\frac{C_{Q}}{\mu^2} = -2\pi R_h^2\;(\beta^2R_h^2+1)\;\frac{(1+R_h^2\beta^2)^2-2\beta^3R_h^2}{(\beta^2R_h^2+1)^3+2\beta^3R_h^2(1-3\beta^2R_h^2)},
\label{cqdymn}
\end{equation}
and is plotted as a function of $\beta$ in Fig.\ref{cqd}.
The RN-like divergence is again present. We have verified that the plot of $C_Q$ tends to $C_Q^{\rm{(Schw)}}$ for large $\beta$, in accordance with Eqn.(\ref{cqdymn}).

\section{Stability according to the Poincar\`e method}

We shall analyze next the stability of the black hole configurations
of the previous sections following the Poincar\`e method.
Let us start by giving a short summary of it, following the presentation in  \citep{Arcioni2004}. 
The Poincar\`e method of stability 
succeeds in showing the existence of unstable modes from the properties of the equilibrium sequence alone, without the need of analyzing the 
Hessian of the system. 
Let ${\cal M}$ be the configuration space of the system under scrutiny, and $X$ a point in ${\cal M}$. The set of independent thermodynamical variables that specify the ensemble shall be denoted by $\{\mu^i\}$, and $S$ will be the corresponding Massieu function. 
The equilibrium states are defined as the points in ${\cal M}$ that are extrema of $S$ under displacements $dX$ for which $d\mu^i = 0$. At any such point, the conjugate thermodynamical variables ${\beta_i}$ are defined by 
\begin{equation}
dS = \beta_id\mu^i
\label{ds}
\end{equation}
for all $dX$. The set of equilibrium points form a submanifold in ${\cal M}$, referred to as ${\cal M}_{eq}$, each point of which can be labelled by the corresponding set of values $\{\mu^i\}$. The maximum entropy postulate states that unconstrained locally stable equilibrium points take place at points in ${\cal M}_{eq}$ in which $S$ is a local maximum with respect to arbitrary variations $dX$ in ${\cal M}$ that preserve $d\mu^i=0$. The fundamental relation $S_{eq} = S(\mu^i)$ can be obtained from Eq.(\ref{ds}), and then 
$$
\beta_i(\mu^i)=\frac{\partial S_{eq}}{\partial \mu^i}
$$
are the equations of state, which give the value of the conjugate variables at equilibrium. 

The starting point of the Poincar\`e method is the fact that the maximum entropy  postulate refers to the behaviour of 
$S$ along off-equilibrium states, not about its variation on ${\cal M}_{eq}$. In other words, the function $S_{eq}$ is not related to local stability without assuming additivity. To inquiry about local stability, an extended Massieu function $\hat S$ is needed, such that it gives the behaviour of $S$ near the equilibrium states. Although $\hat S$ is generally unknown, information about it (and hence about stability) can be obtained with the Poincar\`e method by using only the equilibrium equations of state. We shall give here the recipe that is to be used to identify changes in stability (for the proof of the method see \citep{Katz1978, Parentani1995, Sorkin1981, Sorkin1982}):
a change of stability happens when the plot of a conjugacy diagram $\beta_a(\mu^a)$ (for fixed $a$) along the equilibrium series has a vertical tangent (thus showing a turning point). In other words, a turning point necessarily implies instability, without any other assumptions 
\footnote{Actually, there is another instance that implies a change of stability, namely the existence of a bifurcation point (\textit{i.e.} the crossing point of two sequences of equilibria).}. 

Next we shall apply the turning point method
to the
three static and spherically symmetric black hole solutions with a nonlinear electromagnetic field as a source presented above. 
%
%
In the solutions under scrutiny there are two pairs of conjugate parameters that are of interest regarding stability. These are $(M,\beta_M)$ and $(Q,\beta_Q)$. From the first law of black hole mechanics, 
\begin{equation}
dA = \frac{8\pi}{\kappa_h}dM-8\pi \frac{\Phi_h}{\kappa_h} dQ,
\end{equation} 
where 
the electric potential $\Phi$ at the horizon
is given by
$$
\Phi_h = Q\int_{r_h}^{\infty} \frac{dr}{r^2{\cal L}_F},
$$
we see that $\beta_M = \frac{8\pi}{\kappa_h}$ and $\beta_Q = -8\pi \frac{\Phi_h}{\kappa_h}$. 
It follows from these expressions that the corresponding plots should display divergencies in the case of an extremal black hole
(for which $T=0$ \footnote{See however the discussion in \citep{Liberati2000}.}). We shall see below that this is the case.  

We shall try first to keep
the stability analysis general. Hence, 
no commitment to a specific form of ${\cal L}$ will be adopted. In the case of $(M,\beta_M)$, although we do not have the fundamental relation, the symmetries of the solution allow the calculation of $\frac{\partial M}{\partial \beta_M}$ and of the second derivative for an arbitrary Lagrangian, much in the same way as the $C_Q$ was calculated. 
The results are:  
\begin{equation}
\frac{\partial M}{\partial \beta_M}=\frac{r_h^3}{4\pi}\frac{\kappa_h^3}{1+r_h^2T^0_0(r_h)+r_0^3T^{\prime 0}_0(r_h)},
\label{stab}
\end{equation}
\begin{equation}
\frac{\partial^2 M}{\partial \beta_M^2}  =  \frac{r_h^3\kappa_h^4}{32\pi^2}\;
\frac{
\left[-4r_h^2T^0_0(r_h)(2+r_h^2T^0_0(r_h))
-r_h^3T^{\prime 0}_0(r_h)(7+5r_h^2T^0_0(r_h)+3r_{h}^{3}T^{\prime 0}_{0}(r_h))+r_{h}^{4}
T^{\prime\prime 0}_{0}(r_h)
(r_{h}^{2}T^{0}_{0}(r_h)-1)\right]}{(1+r_h^2T^0_0(r_h)+r_h^3T^{\prime 0}_0(r_h))^3}.
\label{stab2}
\end{equation}
From these expressions we see that the plot of $M$ against $\beta_M$  
cannot have extrema for $\kappa_h \neq 0$. Hence, there are no turning points in the 
($\beta_M, M)$ plane (we shall see below that the plots for the Born-Infeld black hole confirm this assertion). It is also seen that $\frac{\partial\beta_M}{\partial M}$ diverges for the extremal case. Note that these are general statements, in the sense that they are valid for any charged black hole with
the assumed symmetries. 

We can proceed in a similar way with the pair of conjugate variables $(Q,\beta_Q)$. 
The charge of the black hole is given by \citep{Rasheed1997}
\beq
Q = -\frac{1}{8\pi}\varoint dS_{\mu\nu}G^{\mu\nu},
\eeq
where 
\beq
G^{\mu\nu} \equiv -\frac{1}{2} \frac{\partial{\cal L}}{\partial F_{\mu\nu}}.
\eeq
The integral may be calculated over any closed 2-surface enclosing the charge
and is independent of the particular surface chosen. 
In the case of an electric charge, and choosing the horizon as the surface, we get
\beq 
Q = r_h\left.({\cal L}_{F}F^{tr})\right|_h,
\eeq 
which follows also from Eqn.(\ref{maxwell}). Since the calculation of $\beta_Q = -8\pi \frac{\Phi_h}{\kappa_h}$ is carried out at fixed $M$, the rhs is a function of $r_h$. It follows that
\beq
\frac{\partial Q}{\partial \beta_Q} = \frac{\partial Q}{\partial r_h}\frac{\partial r_h}{\partial \beta_Q}.
\eeq
Hence,
\beq
\frac{\partial Q}{\partial \beta_Q} = \frac{\kappa_h^2}{8\pi}\;\frac{
\frac{\partial Q}{\partial r_h}}
{\Phi_h\frac{\partial\kappa}{\partial r_h}-\kappa_h 
\frac{\partial \Phi_h}{\partial r_h}}.
\eeq
To have a turning point, it is necessary that $\frac{\partial Q}{\partial r_h} = 0$. In the case of the R-N black hole, $Q = \sqrt{M^2-(r_h-M)^2}$, and the necessary condition is satisfied only in the extremal case, $r_h = M$. To examine other black hole solutions, the equation $\frac{\partial Q}{\partial r_h} = 0$ should be solved numerically, and the sign of the second derivative should be calculated at the zero of the first derivative. In the next sections we will take an alternative road, which consists in directly plotting $\beta_Q$ 
for the black hole solutions presented in the previous sections. 
As in the previous pair of conjugate variables, $\frac{\partial\beta_Q}{\partial Q}$ diverges for the extremal case.

\subsubsection{Born-Infeld black hole}

\begin{figure}
\centering
\mbox{\subfigure{\includegraphics[width=2in]{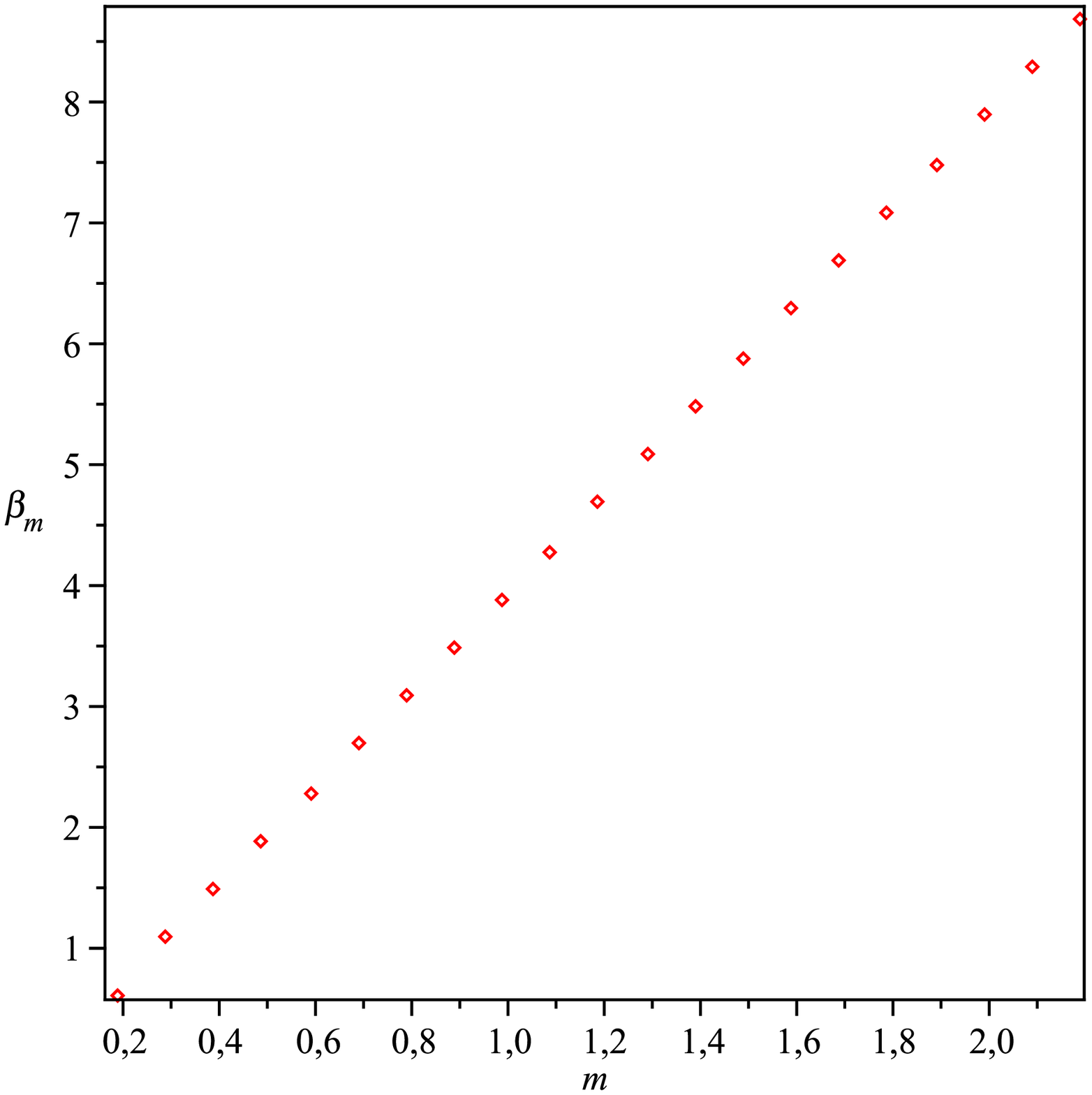}
\quad
\subfigure{\includegraphics[width=2in]{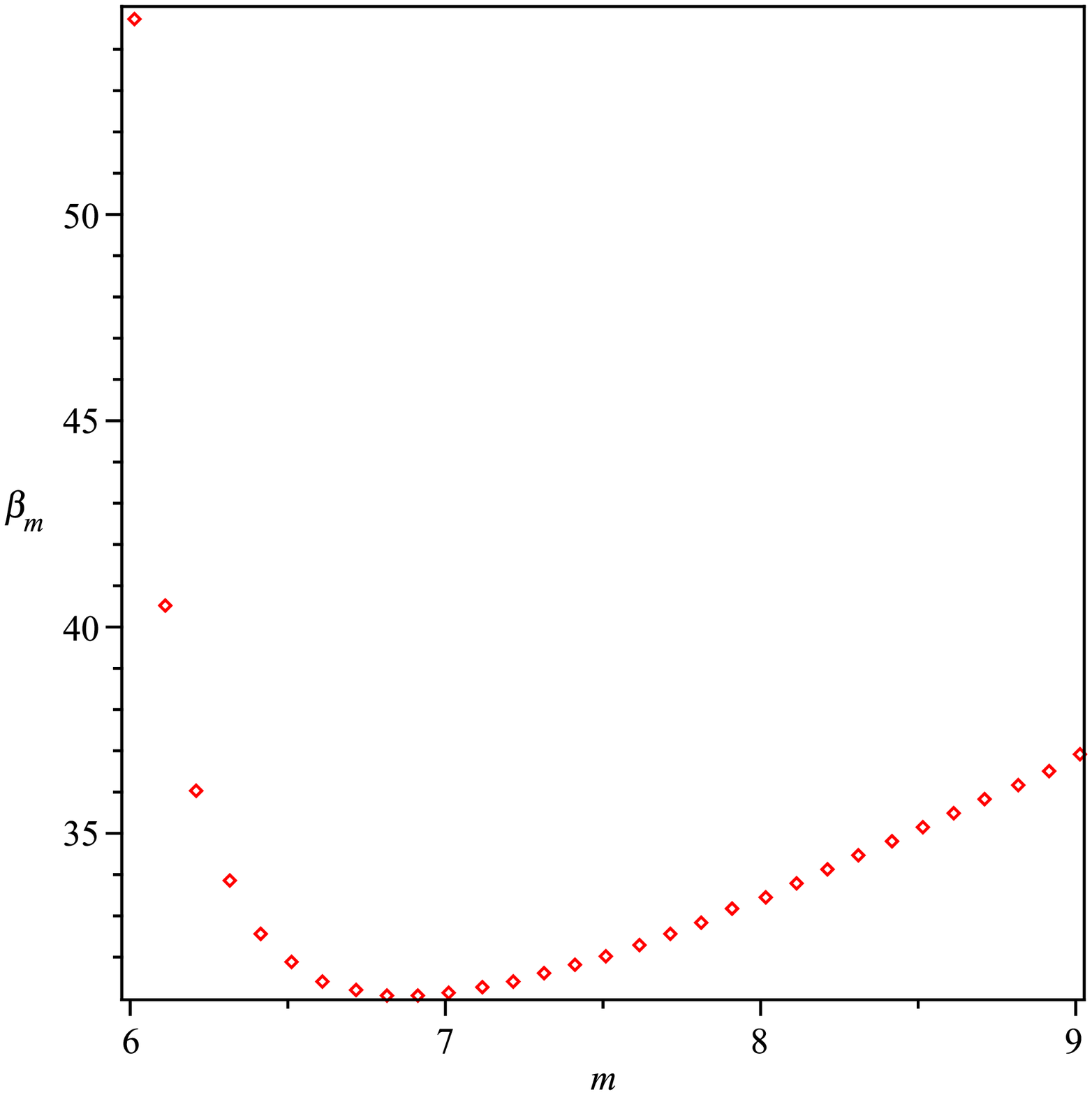}
\quad
\subfigure{\includegraphics[width=2in]{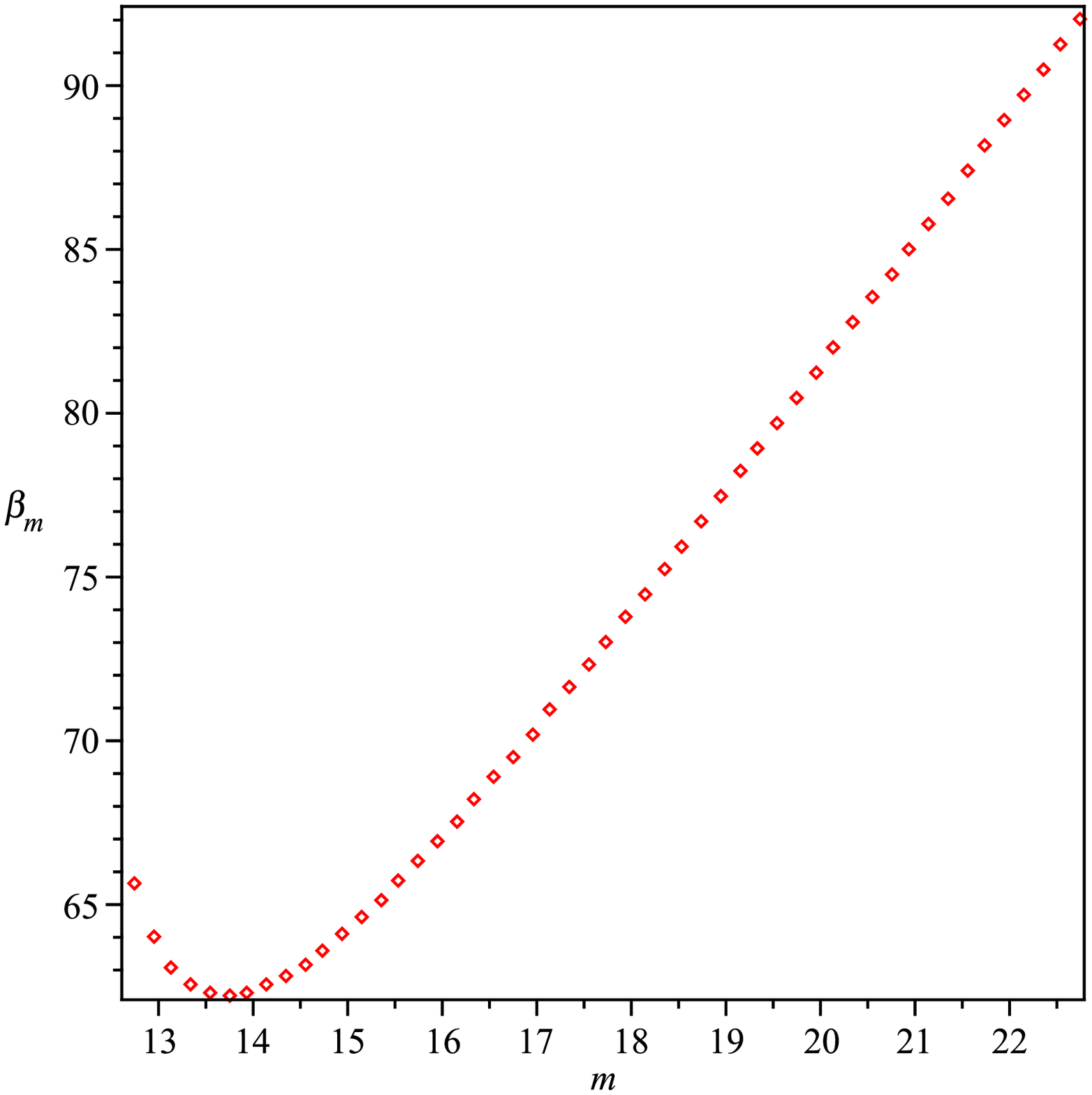}}}}}
\caption{Variation of $\beta_m$ with $m$ at fixed $q=0.2,3,10$, from left to right for the Born-Infeld black hole. The plots do not exhibit turning points.}
\label{qfija}
\end{figure}

We have analitically shown above that there are no turning points in the $(M,\beta_M)$ plane
for any ${\cal L}(F)$. The plots in Fig.\ref{qfija} for the Born-Infeld solution 
are in agreement with this assertion, and display the divergence mentioned above for the extremal black hole.
For the plot of $\beta_Q = -8\pi \frac{\Phi_h}{\kappa_h}$ as a function of $Q$, the electric potential and the surface gravity are needed. They are given by
$$
\Phi_h = q\int_{R_h}^{\infty} \frac{dx}{\sqrt{q^2+x^4}},
$$
\begin{equation}
\frac{\kappa_h}{b}=\frac{1}{ 2 R_h}\left[1+2R_h^2\left(1-
\sqrt{1+\frac{q^2}{R_h^4}}\right)\right]
\label{kappa2}
\end{equation}
Fig.{\ref{mfija} shows that there are no turning points in the 
$\beta_Q$ vs $Q$ plot. The divergence corresponds to the extreme case, given by 
$q^2=\frac{1}{4}+R_h^2$ which, from Eq.(\ref{kappa})
is also where $\kappa_h = 0$.

\begin{figure}
\centering
\mbox{\subfigure{\includegraphics[width=2in]{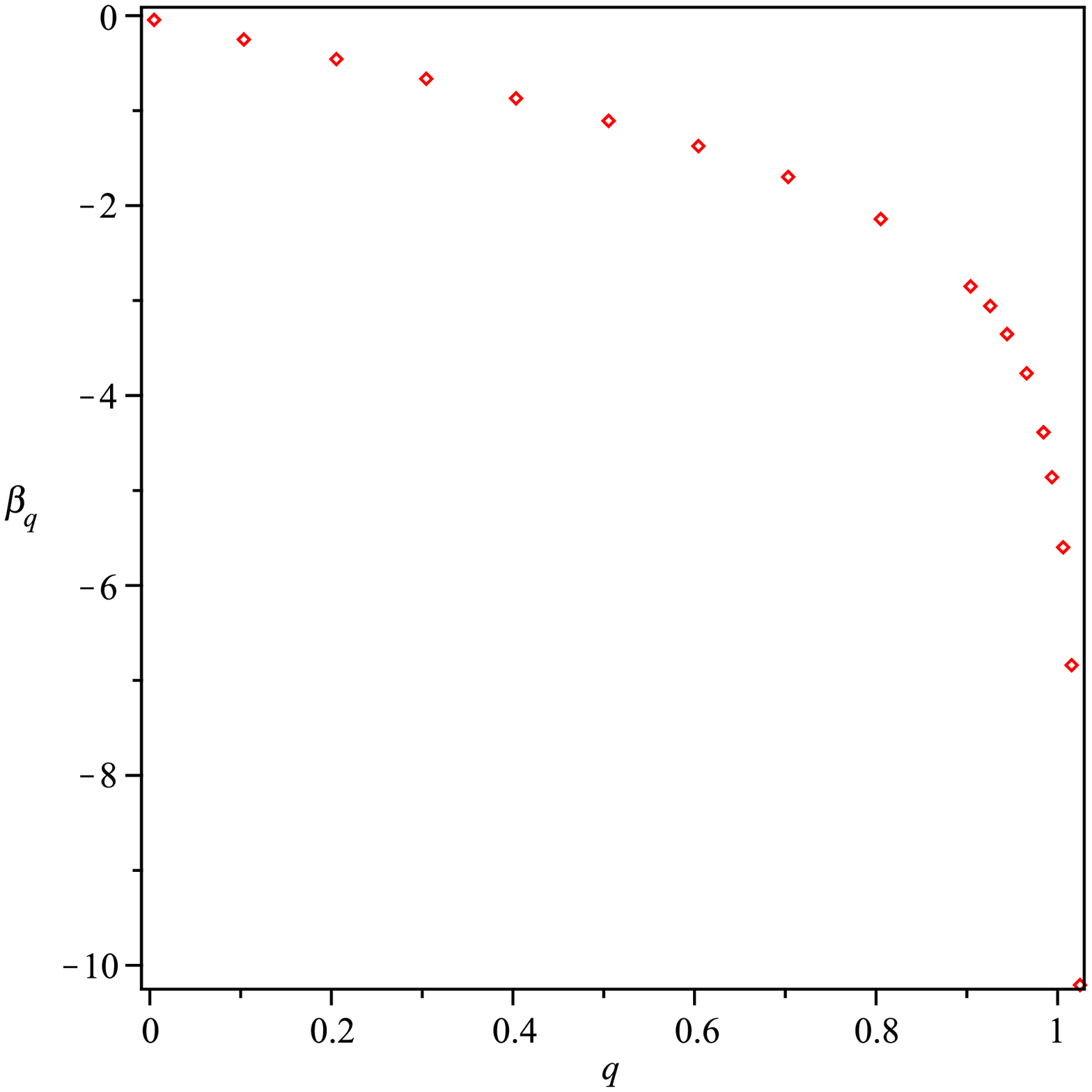}
\quad
\subfigure{\includegraphics[width=2in]{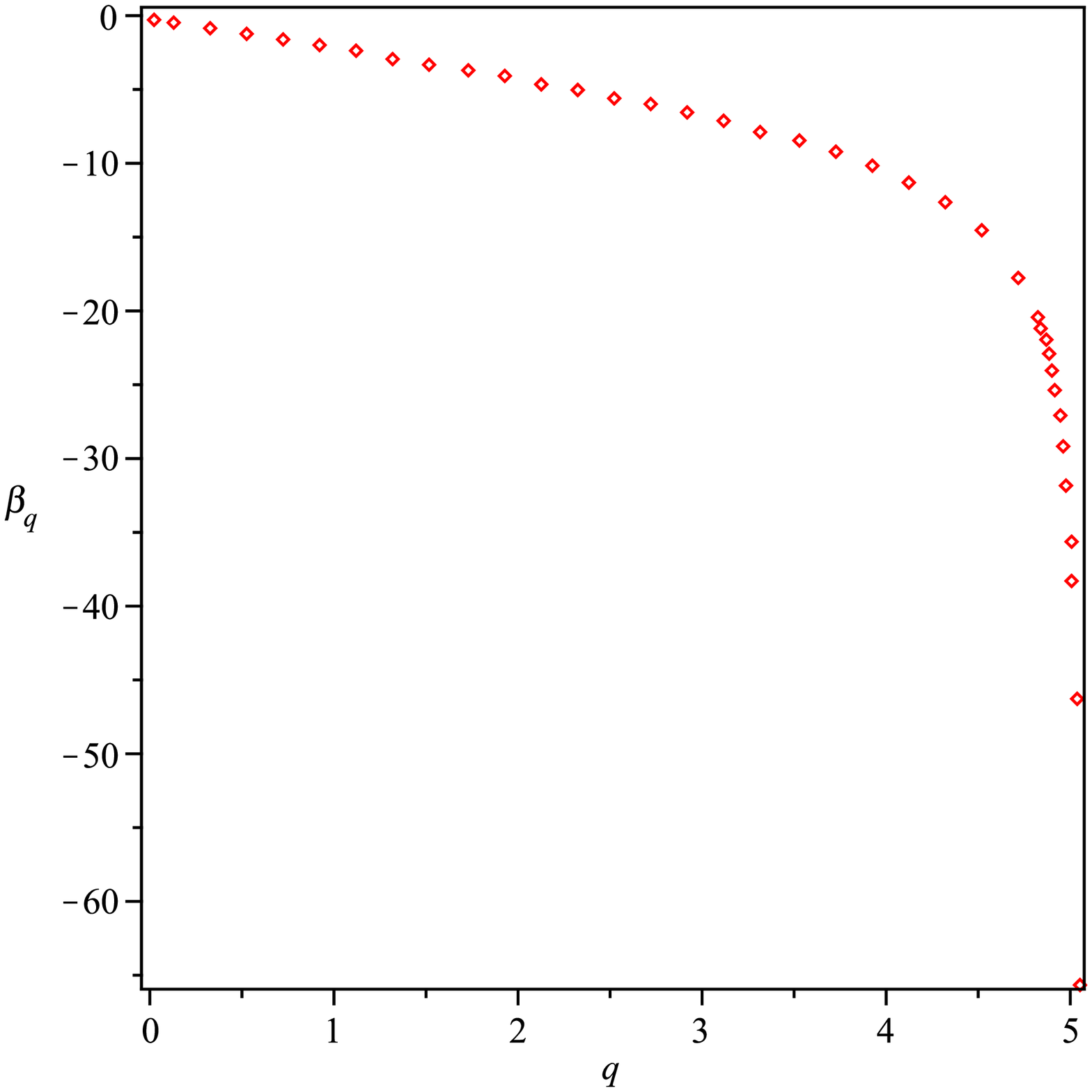}
\quad
\subfigure{\includegraphics[width=2in]{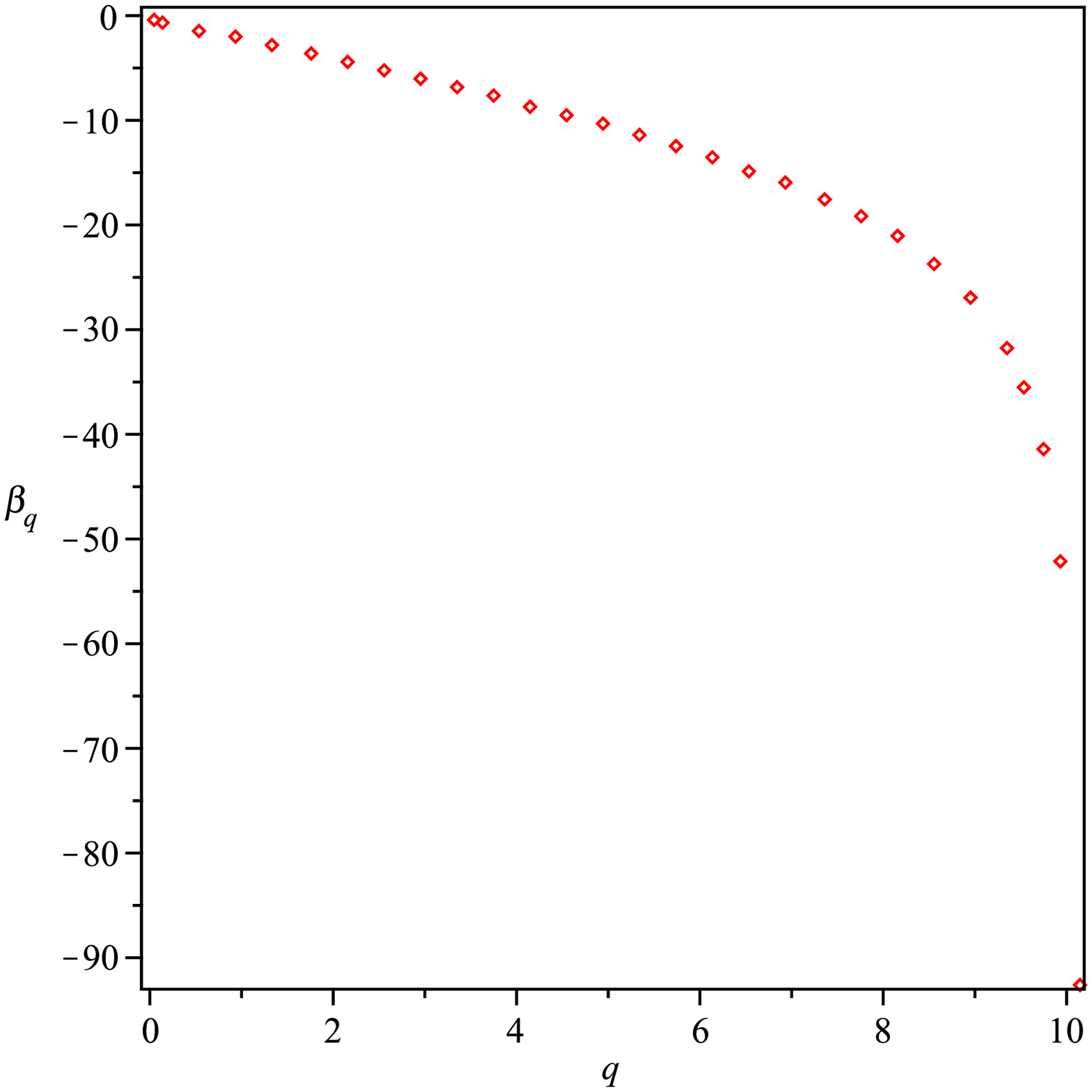}}}}}
\caption{The plots show the variation of $\beta_q$ with the charge, for $m=1,5,10$, from left to right.}
\label{mfija}
\end{figure}
Hence, it follows from the Poincar\`e method that no changes of stability are possible in this solution. 
Since this geometry reduces in the case $Q = 0$ to Schwarschild's solution (which is stable \citep{Vishveshwara1970}), 
our result indicates that the solution is stable
(if no bifurcations are present). 
%

%
We shall present next the plots of $\beta_Q$ as a function of $Q$ for the two nonsingular solutions mentioned above. 

\subsubsection{Bronnikov solution}

As discussed in the begining of this section, 
we need only to examine the behaviour in terms of $Q$
of the function $\beta_Q$. The necessary quantities 
for this calculation are the magnetic potential and the surface gravity, given by
\beq
\psi_h = \frac{1}{\xi}\left\{\frac{1}{4R_h}{\rm sech}^2\left(\frac{1}{2R_h\xi^2}\right)
+\frac 3 2 \xi^2 \tanh\left(\frac{1}{2R_h\xi^2}\right)\right\},
\eeq
\beq
M\kappa_h = \frac{1}{2R_h}\left\{1-\frac{1}{\xi^2 R_h^2}\cosh^{-2}\left(\frac{1}{2R_h\xi^2}
\right)\right\}.
\eeq
The plot of $\beta_Q$ as a function of $Q$ shows no turning points, and the divergence corresponding to the extremal case. Since 
this solution reduces to that of Schwarzschild for large values of $\xi$,
it follows that it is stable (if no 
bifurcations are present).
\begin{figure}[h]
       \centering  
       \includegraphics[width=0.5\textwidth]{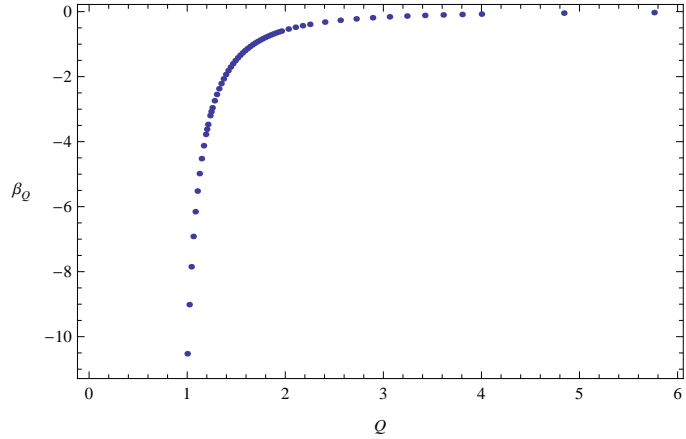}
       \caption{Plot of the conjugate parameters for the Bronnikov solution.}
       \label{pb}
\end{figure}

\subsubsection{Dymnikova solution}
The relevant quantities for the examination of the function $\beta_Q$ are the electric potential and the surface gravity, respectively given by
\beq
\phi_h = \frac{3\sqrt {2\beta}}{8}\left\{\frac\pi 2 -R_h+\frac{2}{3}\beta^3 \frac{R_h^3}{(\beta^2 R_h^2+1)^2}\right\},
\eeq
\beq
\mu\kappa_h = \frac{1}{2R_h}\left(1-2\beta^3\frac{R_h^2}{(\beta^2 R_h^2+1)^2}\right).
\eeq
The dependence of $\beta_Q$ with $Q$ (displaying the divergence for the extremal black hole) is shown in 
Fig.\ref{mfijadymni}.
\begin{figure}[h]
       \centering  
       \includegraphics[width=0.5\textwidth]{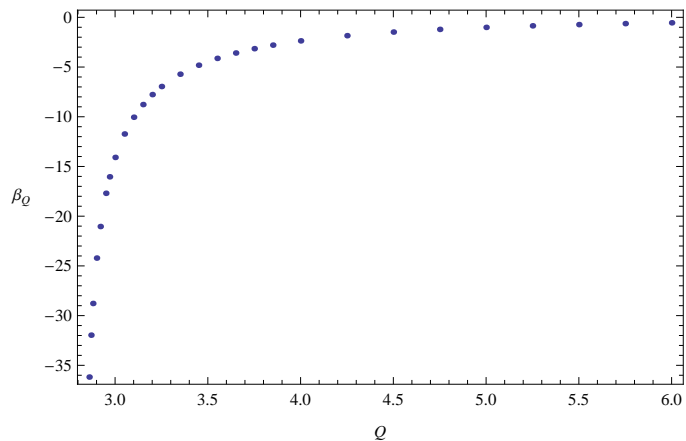}
       \caption{The plot shows the variation of $\beta_Q$ with $Q$ for the Dymnikova solution.}
      \label{mfijadymni}
\end{figure}
As in the previous case, 
the solution is stable (if no 
bifurcations are present), since it reduces to the Schwarzschild black hole for 
large $\beta$.

\section{Comparison of the different methods} 

In this section we compare the results obtained via thermodynamics with 
those that follow from the dynamics of gravitational perturbations
\footnote{Let us remark that the stability of the solutions examined here was inspected in \citep{Breton2005}using the 
Ashtekar-Corichi-Sudarsky (ACS) conjecture. 
In the present work, the situation is different from the one considered in 
\citep{Breton2005} in the boundary conditions. 
The range we considered for the dynamical stability test is the domain of outer communication (DOC), $r \ge r_{h}$, whereas the solitonic solution is not considered at all. In other words, 
we are analyzing completely isolated black holes, otherwise the microcanonical ensemble would not be appropriate for  the thermodynamical analysis, and therefore, the results in \cite{Breton2005} do not correspond to the situation analyzed in this paper.
}.
We have shown that in the three cases examined there is a divergence in the plots of the specific heat at constant charge, as in the RN solution. 
While this divergence has been taken as an indication of thermodynamical instability, our results obtained using the Poincar\`e method showed that
the 
Born-Infeld, Bronnikov, and Dymnikova black holes
are stable, assuming in all cases that no bifurcations are present.

The results obtained via thermodynamics can be compared to those coming 
from a dynamical analysis. 
The dynamical stability of the Born-Infeld black hole under gravitational perturbations has been established in \citep{Fernando2004}.
While we are not aware of any calculation about the dynamical stability
of the two the regular solutions considered here, 
a general analysis of the dynamical stability of SSS black holes with NLEM as a source has been 
presented in \citep{Moreno2002}. As a result, sufficient conditions for dynamical stability were given. 
In the magnetic case they establish that, if 
${\cal L}>0$, ${\cal L}_y>0$, ${\cal L}_{yy}>0$, and $3{\cal L}_y>yg_{00}{\cal L}_{yy}$, the solution is stable, where $y=\sqrt{Q_mF/2}$. All these conditions are satisfied by Bronnikov's solution in the allowed range of parameters, hence it is dynamically stable.

For the electric case, it is convenient to work in the so-called P frame
of nonlinear electrodynamics (see for instance \citep{Salazar1987}). In this case, the sufficient
stability conditions given in \citep{Moreno2002} read
${\cal H}<0$, ${\cal H}_x<0$, ${\cal H}_{xx}<0$, and $3{\cal H}_x\leq x g_{00}{\cal H}_{xx}$, where $x = \sqrt{-2Q_e^2P}$. For Dymnikova's solution, 
$$
{\cal H} = \frac{P}{(1+\alpha\sqrt{-P})^2},$$ and it follows that the solution is dynamically stable when the allowed range of parameters is considered. 

\section{Conclusions}

We derived an expression for the specific heat at constant charge valid for any charged SSS black hole. The $C_Q$ displays, in the three cases examined, a divergence reminiscent of that of the RN solution. 
As a byproduct, 
plots of the radius of the horizon(s) with the relevant parameters of the three solutions considered were obtained. \\
We also studied the thermodynamical stability using the Poincar\`e method. 
Our results for the diagrams of the conjugate variables show that no instability
is present, as is also the case in the RN solution. 
In particular, we have shown that no turning point is possible in the $(\beta_M, M)$ plane for any charged SSS black hole. 

We have also compared our results with those coming from 
studies of dynamical stability. 
Both the Born-Infeld black hole
and the regular solutions are dynamically stable. The absence of turning points for these solutions
might be a hint to the fact that for charged SSS black holes, 
dynamical stability may be equivalent to 
thermodynamical stability \footnote{The same could be said of the RN solution.}, as was shown in \citep{Hollands2012} for vacuum black holes
in General Relativity.  This issue deserves further investigation. 

Regarding the nonsingular solutions, it must be noted that their chararacter seems to change through the addition of gravitational mass, since the coefficient $k$ in Eqn.(\ref{m}) becomes different from zero. As long as the term
$2M_{{\rm grav}}/r$ in this equation remains small, the radius of the horizons (and hence the thermodynamics) will not be very different from the results obtained here. The opposite case, as well as the change in the solution, warrants a closer examination, that will be taken elsewhere.  



\begin{acknowledgments}
N. B. acknowledges partial support from CONACyT, project 166581.
SEPB would like to acknowledge support from FAPERJ, CLAF, UERJ, and CNPQ. 

\end{acknowledgments}

\bibliographystyle{plain} 
\bibliography{bibliography} 

\begin{thebibliography}{10}

\bibitem{Ablu2010}
I.~{Ablu Meitei}, K.~{Yugindro Singh}, T.~{Ibungochouba Singh}, and
  N.~{Ibohal}.
\newblock {Phase transition in the Reissner-Nordstr{\"o}m black hole}.
\newblock {\em \apss}, 327:67--69, May 2010.

\bibitem{Arcioni2004}
Giovanni Arcioni and Ernesto Lozano-Tellechea.
\newblock {Stability and critical phenomena of black holes and black rings}.
\newblock {\em Phys.Rev.}, D72:104021, 2005.

\bibitem{Ayon1998}
Eloy Ayon-Beato and Alberto Garcia.
\newblock {Regular black hole in general relativity coupled to nonlinear
  electrodynamics}.
\newblock {\em Phys.Rev.Lett.}, 80:5056--5059, 1998.

\bibitem{Ayon1999}
Eloy Ayon-Beato and Alberto Garcia.
\newblock {New regular black hole solution from nonlinear electrodynamics}.
\newblock {\em Phys.Lett.}, B464:25, 1999.

\bibitem{Baldovin2000}
F.~Baldovin, M.~Novello, Santiago~E. Perez~Bergliaffa, and J.M. Salim.
\newblock {A Nongravitational wormhole}.
\newblock {\em Class.Quant.Grav.}, 17:3265--3276, 2000.

\bibitem{Bekenstein1998}
J.~D. {Bekenstein}.
\newblock {Black Holes: Classical Properties, Thermodynamics and Heuristic
  Quantization}.
\newblock {\em arXiv:gr-qc/9808028}, August 1998.

\bibitem{Bekenstein1973}
Jacob~D. Bekenstein.
\newblock {Black holes and entropy}.
\newblock {\em Phys.Rev.}, D7:2333--2346, 1973.

\bibitem{Breton2002}
N.~{Bret{\'o}n}.
\newblock {Geodesic structure of the Born-Infeld black hole}.
\newblock {\em Classical and Quantum Gravity}, 19:601--612, February 2002.

\bibitem{Breton2005}
N.~{Bret{\'o}n}.
\newblock {Stability of nonlinear magnetic black holes}.
\newblock {\em \prd}, 72(4):044015, August 2005.

\bibitem{Breton2003}
Nora Breton.
\newblock {Born-Infeld black hole in the isolated horizon framework}.
\newblock {\em Phys.Rev.}, D67:124004, 2003.

\bibitem{Breton2004}
Nora Breton.
\newblock {Smarr's formula for black holes with non-linear electrodynamics}.
\newblock {\em Gen.Rel.Grav.}, 37:643--650, 2005.

\bibitem{Bronnikov2000}
Kirill~A. Bronnikov.
\newblock {Regular magnetic black holes and monopoles from nonlinear
  electrodynamics}.
\newblock {\em Phys.Rev.}, D63:044005, 2001.

\bibitem{Chemissany2008}
Wissam~A. Chemissany, Mees de~Roo, and Sudhakar Panda.
\newblock {Thermodynamics of Born-Infeld Black Holes}.
\newblock {\em Class.Quant.Grav.}, 25:225009, 2008.

\bibitem{Davies1977}
P.~C.~W. {Davies}.
\newblock {The thermodynamic theory of black holes}.
\newblock {\em Royal Society of London Proceedings Series A}, 353:499--521,
  April 1977.

\bibitem{Diaz2009}
J.~Diaz-Alonso and D.~Rubiera-Garcia.
\newblock {Electrostatic spherically symmetric configurations in gravitating
  nonlinear electrodynamics}.
\newblock {\em Phys.Rev.}, D81:064021, 2010.

\bibitem{Diaz2010}
Joaquin Diaz-Alonso and Diego Rubiera-Garcia.
\newblock {Asymptotically anomalous black hole configurations in gravitating
  nonlinear electrodynamics}.
\newblock {\em Phys.Rev.}, D82:085024, 2010.

\bibitem{Dymnikova2004}
Irina Dymnikova.
\newblock {Regular electrically charged structures in nonlinear electrodynamics
  coupled to general relativity}.
\newblock {\em Class.Quant.Grav.}, 21:4417--4429, 2004.

\bibitem{Fernando2004}
Sharmanthie Fernando.
\newblock {Gravitational perturbation and quasi-normal modes of charged black
  holes in Einstein-Born-Infeld gravity}.
\newblock {\em Gen.Rel.Grav.}, 37:585--604, 2005.

\bibitem{Flachi2012}
Antonino Flachi and Jose'~P.S. Lemos.
\newblock {Quasinormal modes of regular black holes}.
\newblock {\em Phys.Rev.}, D87:024034, 2013.

\bibitem{Garcia2012}
Alberto Garcia, Eva Hackmann, Claus Lammerzahl, and Alfredo Macias.
\newblock {No-hair conjecture for Einstein-Plebanski nonlinear electrodynamics
  static black holes}.
\newblock {\em Phys.Rev.}, D86:024037, 2012.

\bibitem{Garcia1984}
A.~{Garc{\'{\i}}a D.}, H.~{Salazar I.}, and J.~F. {Pleba{\'n}ski}.
\newblock {Type-D solutions of the Einstein and Born-Infeld
  nonlinear-electrodynamics equations}.
\newblock {\em Nuovo Cimento B Serie}, 84:65--90, November 1984.

\bibitem{Green2013}
Stephen~R. Green, Joshua~S. Schiffrin, and Robert~M. Wald.
\newblock {Dynamic and Thermodynamic Stability of Relativistic, Perfect Fluid
  Stars}.
\newblock {\em Class.Quant.Grav.}, 31:035023, 2014.

\bibitem{Gunasekaran2012}
Sharmila Gunasekaran, Robert~B. Mann, and David Kubiznak.
\newblock {Extended phase space thermodynamics for charged and rotating black
  holes and Born-Infeld vacuum polarization}.
\newblock {\em JHEP}, 1211:110, 2012.

\bibitem{Hassaine2008}
Mokhtar Hassaine and Cristian Martinez.
\newblock {Higher-dimensional charged black holes solutions with a nonlinear
  electrodynamics source}.
\newblock {\em Class.Quant.Grav.}, 25:195023, 2008.

\bibitem{Hawking1974}
S.W. Hawking.
\newblock {Black hole explosions}.
\newblock {\em Nature}, 248:30--31, 1974.

\bibitem{Hollands2012}
S.~{Hollands} and R.~M. {Wald}.
\newblock {Stability of Black Holes and Black Branes}.
\newblock {\em Communications in Mathematical Physics}, December 2012.

\bibitem{Kaburaki1993}
O.~{Kaburaki}, I.~{Okamoto}, and J.~{Katz}.
\newblock {Thermodynamic stability of Kerr black holes}.
\newblock {\em \prd}, 47:2234--2241, March 1993.

\bibitem{Katz1978}
J.~{Katz}.
\newblock {On the number of unstable modes of an equilibrium}.
\newblock {\em \mnras}, 183:765--770, June 1978.

\bibitem{Katz1993}
J.~{Katz}, I.~{Okamoto}, and O.~{Kaburaki}.
\newblock {Thermodynamic stability of pure black holes}.
\newblock {\em Classical and Quantum Gravity}, 10:1323--1339, July 1993.

\bibitem{Liberati2000}
Stefano Liberati, Tony Rothman, and Sebastiano Sonego.
\newblock {Extremal black holes and the limits of the third law}.
\newblock {\em Int.J.Mod.Phys.}, D10:33--40, 2001.

\bibitem{Lousto1993}
C.~O. {Lousto}.
\newblock {The fourth law of black-hole thermodynamics}.
\newblock {\em Nuclear Physics B}, 410:155--172, December 1993.

\bibitem{Matyjasek2000}
Jerzy Matyjasek.
\newblock {Vacuum polarization of massive scalar fields in the space-time of
  the electrically charged nonlinear black hole}.
\newblock {\em Phys.Rev.}, D63:084004, 2001.

\bibitem{Moncrief1974a}
V.~{Moncrief}.
\newblock {Odd-parity stability of a Reissner-Nordstr{\"o}m black hole}.
\newblock {\em \prd}, 9:2707--2709, May 1974.

\bibitem{Moncrief1974b}
V.~{Moncrief}.
\newblock {Stability of Reissner-Nordstr{\"o}m black holes}.
\newblock {\em \prd}, 10:1057--1059, August 1974.

\bibitem{Moreno2002}
Claudia Moreno and Olivier Sarbach.
\newblock {Stability properties of black holes in selfgravitating nonlinear
  electrodynamics}.
\newblock {\em Phys.Rev.}, D67:024028, 2003.

\bibitem{Note1}
See \protect \citep {Poincare1885} for the original version, and \protect
  \citep {Sorkin1982} and \protect \citep {Schiffrin2013} for updates.

\bibitem{Note10}
Let us remark that the stability of the solutions examined here was inspected
  in \protect \citep {Breton2005}using the Ashtekar-Corichi-Sudarsky (ACS)
  conjecture. In the present work, the situation is different from the one
  considered in \protect \citep {Breton2005} in the boundary conditions. The
  range we considered for the dynamical stability test is the domain of outer
  communication (DOC), $r \ge r_{h}$, whereas the solitonic solution is not
  considered at all. In other words, we are analyzing completely isolated black
  holes, otherwise the microcanonical ensemble would not be appropriate for the
  thermodynamical analysis, and therefore, the results in \cite {Breton2005} do
  not correspond to the situation analyzed in this paper.

\bibitem{Note11}
The same could be said of the RN solution.

\bibitem{Note2}
For different points of view about this discrepancy, see \protect \citep
  {Sokolowski1980,Sorkin1982, Pavon1988, Pavon1991, Katz1993,
  Kaburaki1993,Lousto1993, Arcioni2004, Ablu2010,Parentani1995, Okamoto1995}.

\bibitem{Note3}
The calculation of $C_\Phi $, and of the analogous of the thermal expansion and
  the isotermal compressibility involve Legendre transformations, and the use
  of the Smarr formula, which is not available in this context.

\bibitem{Note4}
The extremal case, in which the two horizons coalesce into one, is given by
  $q^2={\begingroup 1\endgroup \over 4}+R_h^2$ \protect \citep
  {Chemissany2008}.

\bibitem{Note5}
This Lagrangian was inspired in the one presented in \protect \citep
  {Ayon1999}.

\bibitem{Note6}
As shown in \protect \citep {Matyjasek2000}, the location of the horizons can
  be expressed in terms of the Lambert function.

\bibitem{Note7}
This solution evades the no-go theorem presented in \protect \citep
  {Bronnikov2000} because the Lagrangian does not go to the Maxwell's limit at
  the center, see \protect \citep {Dymnikova2004}.

\bibitem{Note8}
Actually, there is another instance that implies a change of stability, namely
  the existence of a bifurcation point (\protect \textit {i.e.} the crossing
  point of two sequences of equilibria).

\bibitem{Note9}
See however the discussion in \protect \citep {Liberati2000}.

\bibitem{Okamoto1995}
I.~{Okamoto}, J.~{Katz}, and R.~{Parentani}.
\newblock {A comment on fluctuations and stability limits with application to
  `superheated' black holes}.
\newblock {\em Classical and Quantum Gravity}, 12:443--448, February 1995.

\bibitem{Parentani1995}
R.~{Parentani}, J.~{Katz}, and I.~{Okamoto}.
\newblock {Thermodynamics of a black hole in a cavity}.
\newblock {\em Classical and Quantum Gravity}, 12:1663--1684, July 1995.

\bibitem{Pavon1991}
D.~{Pav{\'o}n}.
\newblock {Phase transition in Reissner-Nordstr{\"o}m black holes}.
\newblock {\em \prd}, 43:2495--2497, April 1991.

\bibitem{Pavon1988}
D.~{Pav{\'o}n} and J.~M. {Rub{\'{\i}}}.
\newblock {Nonequilibrium thermodynamic fluctuations of black holes}.
\newblock {\em \prd}, 37:2052--2058, April 1988.

\bibitem{Pellicer1969}
R.~Pellicer and R.J. Torrence.
\newblock {Nonlinear electrodynamics and general relativity}.
\newblock {\em J.Math.Phys.}, 10:1718--1723, 1969.

\bibitem{Poincare1885}
H.~{Poincar\`e}.
\newblock {}.
\newblock {\em Acta Math.}, 7:259, {} 1885.

\bibitem{Rasheed1997}
D.A. Rasheed.
\newblock {Nonlinear electrodynamics: Zeroth and first laws of black hole
  mechanics}.
\newblock 1997.

\bibitem{Regge1957}
T.~{Regge} and J.~A. {Wheeler}.
\newblock {Stability of a Schwarzschild Singularity}.
\newblock {\em Physical Review}, 108:1063--1069, November 1957.

\bibitem{Salazar1987}
I.H. Salazar, A.~Garcia, and J.~Plebanski.
\newblock {Duality Rotations and Type $D$ Solutions to Einstein Equations With
  Nonlinear Electromagnetic Sources}.
\newblock {\em J.Math.Phys.}, 28:2171--2181, 1987.

\bibitem{Schiffrin2013}
Joshua~S. Schiffrin and Robert~M. Wald.
\newblock {Turning Point Instabilities for Relativistic Stars and Black Holes}.
\newblock 2013.

\bibitem{Sokolowski1980}
L.~M. {Sokolowski} and P.~{Mazur}.
\newblock {Second-order phase transitions in black-hole thermodynamics}.
\newblock {\em Journal of Physics A Mathematical General}, 13:1113--1120, March
  1980.

\bibitem{Sorkin1981}
R.~{Sorkin}.
\newblock {A Criterion for the Onset of Instability at a Turning Point}.
\newblock {\em \apj}, 249:254, October 1981.

\bibitem{Sorkin1982}
R.~D. {Sorkin}.
\newblock {A Stability Criterion for Many Parameter Equilibrium Families}.
\newblock {\em \apj}, 257:847, June 1982.

\bibitem{Vishveshwara1970}
C.~V. {Vishveshwara}.
\newblock {Stability of the Schwarzschild Metric}.
\newblock {\em \prd}, 1:2870--2879, May 1970.

\bibitem{Visser1992}
Matt Visser.
\newblock {Dirty black holes: Thermodynamics and horizon structure}.
\newblock {\em Phys.Rev.}, D46:2445--2451, 1992.

\bibitem{Whiting1989}
B.~F. {Whiting}.
\newblock {Mode stability of the Kerr black hole.}
\newblock {\em Journal of Mathematical Physics}, 30:1301--1305, June 1989.

\bibitem{Yajima2000}
Hiroki Yajima and Takashi Tamaki.
\newblock {Black hole solutions in Euler-Heisenberg theory}.
\newblock {\em Phys.Rev.}, D63:064007, 2001.

\end{thebibliography}

\end{document}